\documentclass{jfm}

\usepackage{bm}
\usepackage{graphicx}
\usepackage{newtxtext}
\usepackage{newtxmath}
\usepackage{natbib}
\usepackage{hyperref}
\usepackage{float}
\usepackage{amsmath}
\usepackage{enumerate}   
\usepackage{amssymb}

\usepackage[justification=justified]{caption}
\usepackage{xcolor}
\usepackage{lineno-FLM}
\nolinenumbers

\usepackage{subcaption}
\usepackage{color}

\hypersetup{
    colorlinks = true, 
    urlcolor   = blue, 
    citecolor  = blue,
    urlcolor= blue, 
linkcolor= blue, 
bookmarksopen=true, 
}

\newcommand{\RomanNumeralCaps}[1]

\newcommand{\cE}{\mathcal{E}}

\title{Collapse of statistical equilibrium in large-scale hydroelastic turbulent waves}
\author{Marlone Vernet  \and Eric Falcon\corresp{\email{eric.falcon@u-paris.fr}}}
\affiliation{Université Paris Cité, CNRS, MSC, UMR 7057, F-75 013 Paris, France
}
\begin{document}
\maketitle
\begin{abstract}
At scales larger than the forcing scale, some out-of-equilibrium turbulent systems (such as hydrodynamic turbulence, wave turbulence, and nonlinear optics) exhibit a state of statistical equilibrium where energy is equipartitioned among large-scale modes, in line with the Rayleigh-Jeans spectrum. Key open questions now pertain to either the emergence, decay, collapse, or other nonstationary evolutions from this state. Here, we experimentally investigate the free decay of large-scale hydroelastic turbulent waves, initially in a regime of statistical equilibrium. Using space- and time-resolved measurements, we show that the total energy of these large-scale tensional waves decays as a power law in time. We derive an energy decay law from the theoretical initial equilibrium spectrum and the linear viscous damping, as no net energy flux is carried. Our prediction then shows a good agreement with experimental data over nearly two decades in time, for various initial effective temperatures of the statistical equilibrium state. We further identify the dissipation mechanism and confirm it experimentally. Our approach could be applied to other decaying turbulence systems, with the large scales initially in statistical equilibrium.

\end{abstract}
\section{Introduction}

The possible coexistence of two distinct regimes in out-of-equilibrium (forced and dissipated) turbulent systems, namely, a turbulent cascade at small scales and statistical equilibrium (SE) at large scales, is particularly interesting. It suggests that classical tools from statistical mechanics may be applicable for describing the large-scale behavior of turbulent systems. In this context, ``large scales'' refer to scales larger than the forcing scale, and SE is characterized by energy equipartition among degrees of freedom due to the absence of flux across large scales. Evidence of SE at large scales has been reported in different systems, including three-dimensional (3D) hydrodynamic turbulence~\citep{gorce2022PRL}, capillary wave turbulence~\citep{michel2017}, flexural wave turbulence~\citep{miquel2021PRE}, and hydroelastic wave turbulence~\citep{vernet25}. Classical condensation toward SE has also been reported in optics~\citep{SunNatPhys2012,BaudinPRL2020}. 

While the existence of SE within out-of-equilibrium systems is appealing, many questions remain. In particular, what is the mechanism to reach the SE regime, and how does it grow or decay upon the starting or stopping of external forcing?  How do the regimes of statistical equilibrium (at large scales) and out-of-equilibrium (at small scales) interact, since they are not isolated from each other? These questions, in this context, involve nonstationary processes that are central to understanding how stationary regimes in physics arise and collapse. 

Numerous studies have focused on the decay processes in nonlinear out-of-equilibrium systems, including 3D hydrodynamic turbulence in wind tunnels~\citep{comte1966} and rotating turbulence~\citep{morize2006}, initially forced at a large scale. 
Free decay has also been investigated in two-dimensional (2D) turbulence, in a soap film~\citep{martin1998,cieslik2009}, in fluid layers~\citep{shats2010}, in electron plasma~\citep{bettega2009}, in 3D magnetohydrodynamics turbulence~\citep{bigot2008,beresnyak2008}, and quantum turbulence~\citep{skrbek2012}. In the context of wave turbulence, decay dynamics have been explored numerically for surface gravity waves~\citep{onorato2002,Yokoyama2004,Zhang2022} and capillary waves~\citep{Falkovich1995,pan2015} and experimentally for elastic waves in a thin plate~\citep{miquel2011}, hydroelastic waves~\citep{deike2013JFM}, surface gravity~\citep{Bedard2013,deike2015}, gravity-capillary~\citep{cazaubiel2019} and capillary~\citep{kolmakov2004,deike2012} waves.  

While all these nonstationary studies focused on the free decay of small-scale turbulent cascade (primarily investigating whether the constant energy-flux phenomenology persists), only a few have examined the decay behavior of large scales that initially exhibit a SE regime. In particular, a recent experiment on 3D turbulence, initially forced at a small scale, was able to distinguish between Saffman's and Batchelor's models of freely decaying turbulence, in the early stage and final decay regimes~\citep{gorce2024PRL}.

Hydroleastic waves are of particular interest, as they are deformation modes of an elastic plate in contact with a fluid. These waves have direct applications in oceanography, notably in modeling wave propagation over ice-covered ocean surfaces~\citep{Davys1985,sutherland2016,das2018,Parau2024}. The need for a better understanding of hydroelastic wave dynamics has grown in recent years with the emergence of very large floating structures, such as mobile offshore bases and expansive floating solar farms~\citep{wang2008,yang2024}.

Here, we experimentally investigate how large-scale hydroelastic turbulent waves, initially in SE, freely decay when external small-scale forcing is stopped. {This decay is then understood using a simple model based on the energy budget in Fourier space and on the dissipation mechanism at play.}

\section{Experimental setup}\label{sec:exp}
\begin{figure}
\centering
\includegraphics[width=0.8\columnwidth]{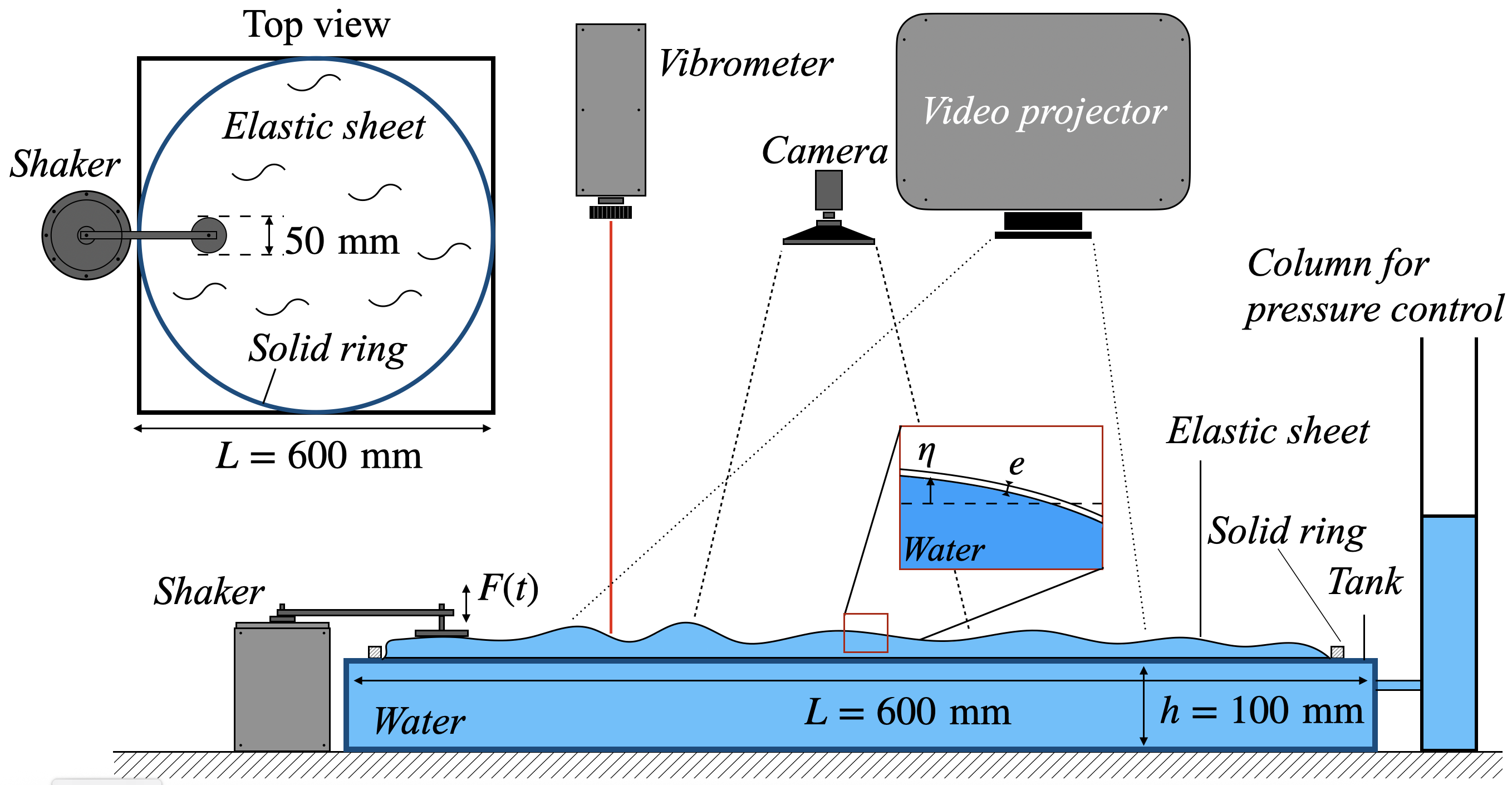}\\
\includegraphics[width=0.6\columnwidth]{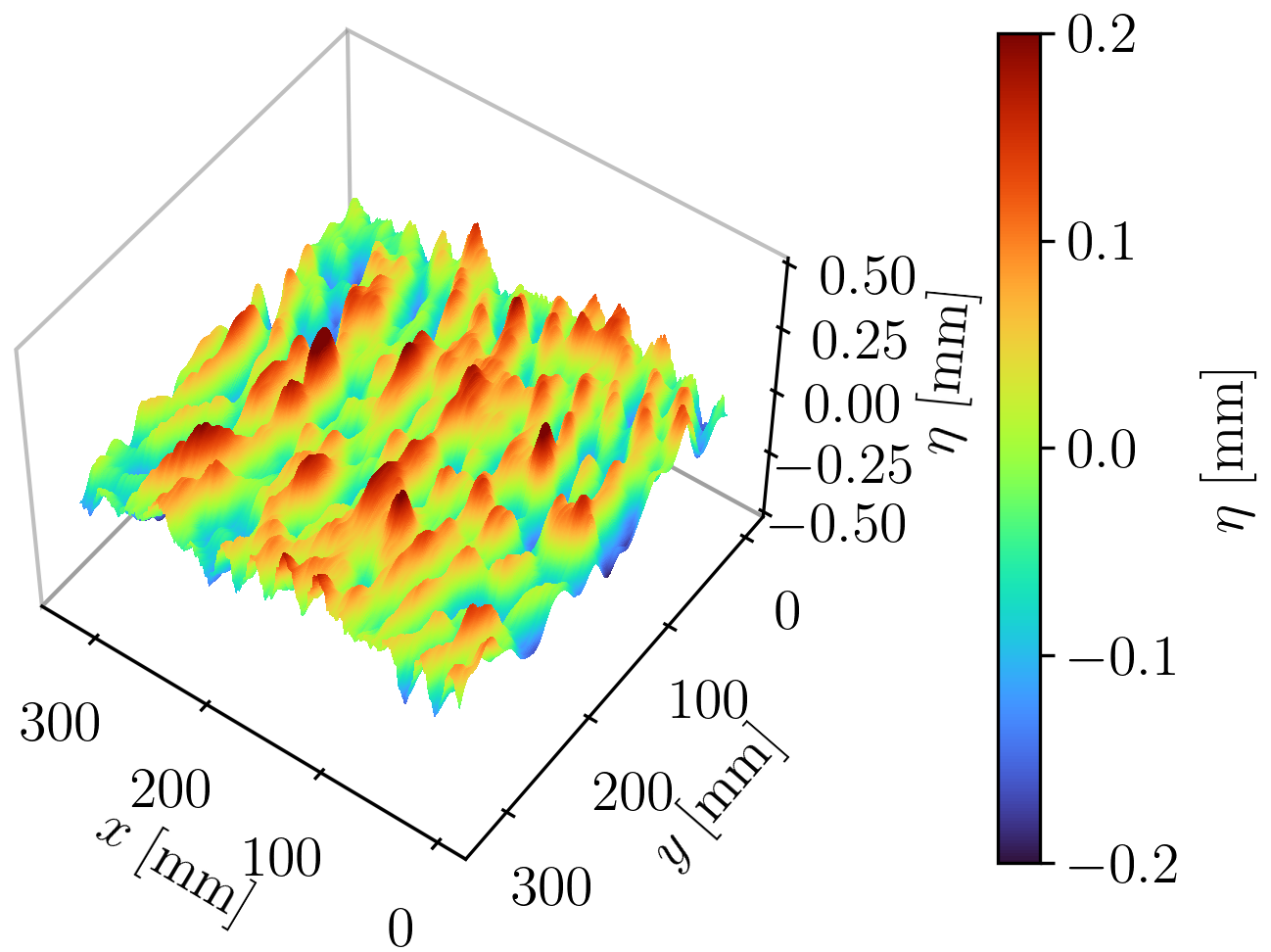}  
\caption{\label{fig:setup}Top: Scheme of the experimental setup {to study hydroelastic waves} (lateral view). Left: top view. Bottom: Typical hydroelastic wave field, $\eta(x,y)$, obtained from Fourier transform profilometry.}
\end{figure}

The experimental setup follows that of~\cite{vernet25} and is shown in figure~\ref{fig:setup}. It consists of a square tank of dimensions $L\times L\times h$ with $L=600$~mm and $h=100$~mm. The tank is fully filled with water {(density $\rho=10^3$~kg\ m$^{-3}$)} and covered by a white elastic sheet made of silicone rubber (Ecoflex 00-30 soft elastomer) of thickness $e=0.5$~mm, density $\rho_s\simeq 10^3$~kg\ m$^{-3}$, Young's modulus $E_Y = 7\times 10^4$~Pa and Poisson's coefficient $\tilde{\nu}\approx 0.5$~\citep{DelorySM2024}. A solid ring is placed on the top of the sheet to apply circular boundary conditions and to enhance multidirectional wave reflections. The tank is connected to a column filled with water to control the imposed hydrostatic pressure in the liquid and the sheet stretching, and thus the applied tension $T\in[3,7]$ N\ m$^{-1}$ to the sheet. Hydroelastic waves are produced by the vertical motion of a disk-shaped wavemaker ($50$~mm in diameter) driven by an electromagnetic shaker (LDS V406) fed with a bandpass-filtered Gaussian random noise signal in the range $f_p\in[50,100]$~Hz. The vertical deformations of the sheet are either measured at a given position, {$\xi(t)$}, with a laser Doppler vibrometer (Polytech OFV5000-505) or are fully resolved in space and time, $\eta(x,y,t)$, using the Fourier transform profilometry (FTP) method~\citep{cobelli2009} using a camera (Basler acA2040) recording at 120 fps the deformations of a fringe pattern projected over the sheet by a full-HD video projector (Epson EH-TW9400). The size of the recorded images, centered in the middle of the sheet, is $\mathcal{L}^2=300 \times 300$~mm$^2$. The sampling frequency of the vibrometer is 2~kHz. The two notations $\xi$ and $\eta$ thus refer to the two experimental methods that yield the sheet vertical deformations. 

The two measurements are not performed simultaneously due to technical reasons. In both cases, measurements are started at $t = 0$ when the wave maker is turned on to capture the statistical equilibrium regime of the large scales (reached in about 1~s). At $t_0$, it is switched off to record the free decay. The same experiment is repeated 60 times, in the vibrometer case, to obtain an ensemble average, thereby increasing statistics, reducing the signal-to-noise ratio, and improving the time-frequency analysis. This one-point measurement is used to perform a time-frequency analysis of the decay lasting $\sim 20$~s. The spatiotemporal measurement gives access to the spatial decay of the Fourier modes for $t\geq t_0$ and the dispersion relation for $t<t_0$.   Typical wave amplitude is [0.1, 1]~mm in the SE regime. 

\section{Results}
\subsection{Hydroelastic waves}\label{sec:LDR}
Hydroelastic waves are deformations of an elastic sheet covering a fluid. They are described by combining two equations, coupling elasticity and hydrodynamics. The Föppl-von K\'arm\'an equation expresses the dynamics of the vertical deformation $\eta(x,y,t)$ of a thin elastic sheet while Bernoulli's theorem yields the pressure field exerted by a perfect fluid on the sheet~\citep{Landau}. Neglecting sheet inertia ($ke\ll \rho/\rho_s$ with $\rho$ the fluid density), the dispersion relation of hydroelastic waves, in a deep water regime ($kh\gg 1$ with fluid depth $h$), reads~\citep{Davys1985,schulkes1987,deike2013JFM} 
\begin{equation}
	\omega^2 = gk+\frac{T}{\rho}k^3 + \frac{B}{\rho}k^5,
	\label{eq:LDR}
\end{equation}
where $\omega$ is the wave angular frequency, $k$ is the wave number modulus, $g$ is the gravity acceleration, $T$ is the externally applied tension, and $B=E_Y e^3 /[12(1-\tilde{\nu}^2)]\simeq 9.7\ 10^{-7}$~{N~m} is the bending modulus. The three terms of the right-hand side of Eq.~\eqref{eq:LDR} correspond to gravity waves, tensional waves, and flexural waves, respectively.
\begin{figure}
\centering
\includegraphics[width=0.65\columnwidth]{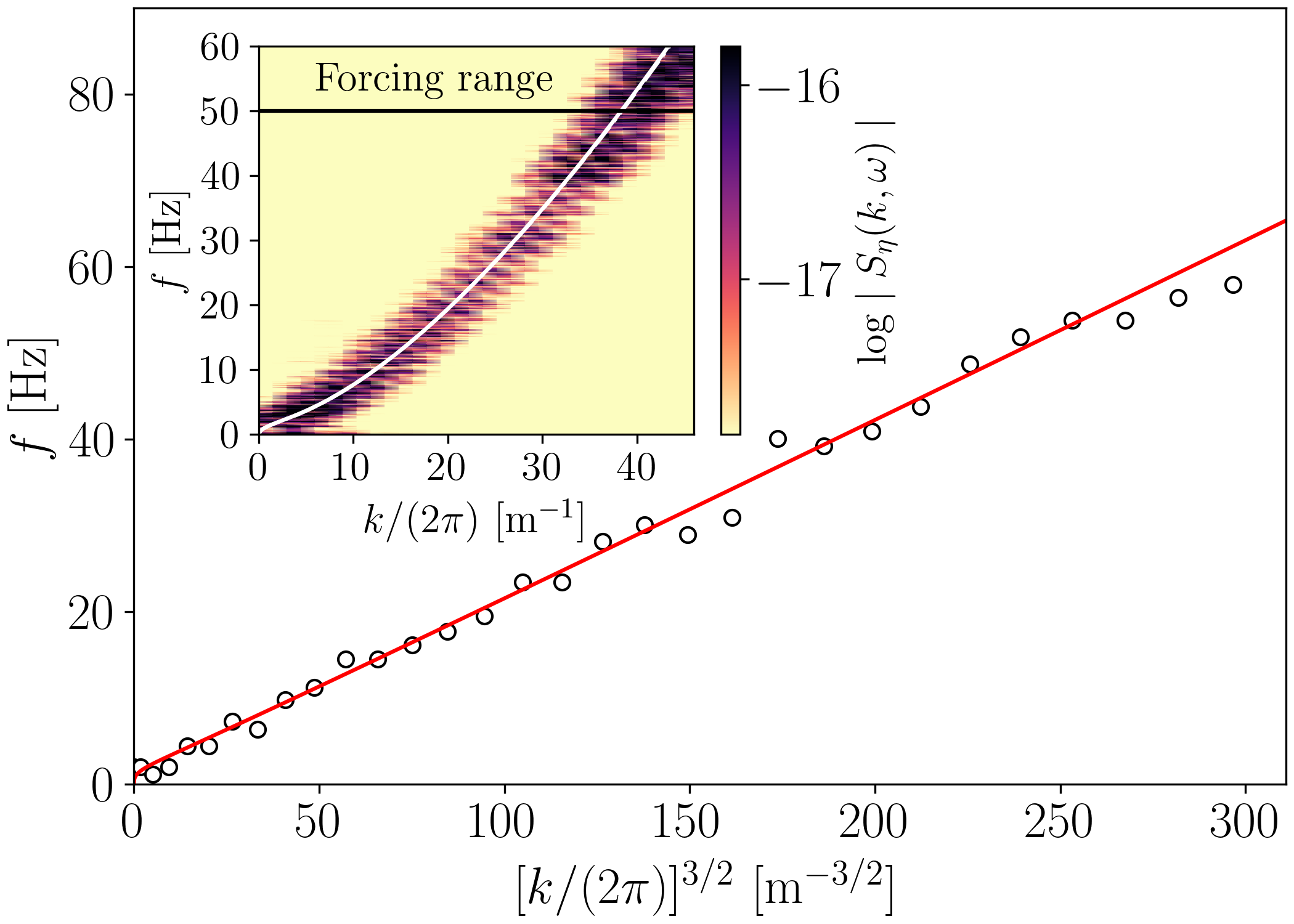}
\caption{\label{fig:LDR} Experimental dispersion relation $f$ versus $[k/(2\pi)]^{3/2}$ (black circle). Small-scale random forcing $f_p\in[50, 100]$~Hz. Red line: polynomial fit of Eq.~(\ref{eq:LDR}) with only one fitting parameter, the tension $T$, which yields $T=6.8$~N~m$^{-1}$. Inset: Spatiotemporal spectrum of the wave height, $S_\eta(k,\omega)$, versus $f\equiv \omega/(2\pi)$ and $k/(2\pi)$. Solid white line: linear dispersion relation of Eq.~(\ref{eq:LDR}). Black line: lower bound [$\min(f_p)$] of the random forcing range.}
\end{figure}

The dispersion relation is reached experimentally from the wave field, $\eta(x,y,t)$, given by the FTP method. We compute the Fourier transform in space {$\check{\eta}(k_x,k_y,t)$} which leads in polar coordinates to $\check{\eta}(k,\varphi,t)$ with $k=(k_x^2+k_y^2)^{1/2}$ and $\varphi\in[0,2\pi[$. The Fourier transform in time of this field gives $\hat{\eta}(k,\varphi,\omega)$ which is used to compute the spatiotemporal spectrum $S_\eta(k,\omega) = \langle \vert \hat{\eta}(k,\varphi,\omega)\vert^2 \rangle_{\varphi}/(\mathcal{L}^2\mathcal{T})$, with $\mathcal{L}$ the window length and $\mathcal{T}=30$~s the acquisition time. The spatiotemporal spectrum, $S_\eta(k,\omega)$, of the wave field, $\eta(x,y,t)$, is shown in the inset of figure~\ref{fig:LDR}. A small-scale random forcing is applied in the range $f_p\in[50, 100]$~Hz. We observe that wave energy is spread over all scales larger than the forcing scales (i.e., $f<f_p$), around the theoretical dispersion relation of Eq.~\eqref{eq:LDR} (white line). The experimental dispersion relation is then inferred from the $S_\eta(k,\omega)$ maximum value for each $k$ as shown in the main figure~\ref{fig:LDR} (black circle). Experimental data exhibit a clear $k^{3/2}$ power law, highlighting the tensional nature of the waves over the $f$ and $k$ available ranges. Experimental data are then fitted by a polynomial fit of Eq.~(\ref{eq:LDR}) to infer the sheet tension $T$. Note that from Eq.~\eqref{eq:LDR}, the crossover between gravity and tension waves is $k_{gt}=\sqrt{\rho g/T}$, and between bending and tension waves, $k_{bt}=\sqrt{T/B}$, which give $k_{gt}/(2\pi)\simeq 6$~m$^{-1}$ and $k_{bt}/(2\pi)\simeq 42{1}$~m$^{-1}$, respectively. Gravity is thus relevant for wavelengths longer than $\sim 16$~cm, whereas bending effects occur at scales much smaller than the forcing scale, thus confirming the tensional nature of the waves in the window of interest.

\subsection{Statistical equilibrium {(Rayleigh-Jean spectrum)}}\label{sec:StatEq}
Statistical equilibrium corresponds to a {stationary state} where energy is equally shared among different modes of a system. For 2D isotropic systems, such equipartition of kinetic energy among Fourier modes $\mathbf{k}$ leads to the theoretical energy spectrum~\citep{zakharov1992}
\begin{align}
	E^{\mathrm{Eq}}(k) = \frac{k_B\theta}{2\pi\rho}k,
\end{align}
where $k\equiv||\mathbf{k}||$, $k_B$ is the Boltzmann constant and $\theta$ is an effective temperature.  Using $E^{\mathrm{Eq}}(k)dk=E^{\mathrm{Eq}}({\omega})d\omega$, and Eq.~\eqref{eq:LDR} considering only tensional waves, the energy spectrum of statistical equilibrium of large-scale hydroelastic tensional waves reads
\begin{align}
	E^\mathrm{Eq}(\omega) = \frac{k_B\theta}{3\pi \rho^{1/3} T^{2/3}}\omega^{1/3}.
	\label{eq:SE_omega}
\end{align}
Note that the energy spectrum is related to the wave amplitude power spectrum, {$S(k)$,} by $E(k) = (T/\rho)k^2S(k)$ and $E(\omega) = (T/\rho)^{1/3}\omega^{4/3}S(\omega)$ for pure tensional waves. Such a statistical equilibrium regime of large-scale hydroelastic tensional waves has been experimentally evidenced by~\cite{vernet25} (see also Sect.~\ref{sec:Decay}). Hereafter, as $\theta$ is no longer discussed, we will use $E_0\equiv k_B\theta$ as the SE energy reference.

\subsection{Free decay from a statistical equilibrium regime}\label{sec:cool}
We now focus on the free decay of a system initially in a statistical equilibrium regime, once the forcing is stopped, to understand how the energy stored in large scales is damped. To do so, we adapt to our case the model of decay of 3D hydrodynamics turbulence~\cite{batchelor1953,Saffman67,gorce2024PRL}.

The energy budget in $\omega$-Fourier space reads
\begin{equation}
	\frac{\partial E(\omega,t)}{\partial t} = T(\omega,t) + I(\omega,t) - D(\omega,t),
	\label{eq:balanceEfull}
\end{equation}
where $E(\omega,t)$ is the energy spectral density, $T(\omega,t)$ is the energy transfer between waves (nonlinear term), $I(\omega,t)$ denotes the injected power, and $D(\omega,t)$ the spectral dissipation (linear term). In the stationary state, dissipation is negligible compared to nonlinearities within the large scales. As soon as the Rayleigh-Jean spectrum is established, the mean energy flux vanishes at large scales. This does not exclude positive or negative fluctuations of the energy flux, which are the counterpart of the fluctuations of the injected power, as observed for the large scales in SE in hydrodynamic turbulence~\citep{gorce2022PRL}.
Once the external forcing is turned off at $t=t_0$, one has $I=0$ and the wave field then freely decays over time, in two stages.\\ (i) First, an initial decay dominated by reminiscent nonlinear processes, which lasts as long as the energy transfer towards small scales occurs in the inertial range~\citep{batchelor1953,comte1966,gorce2024PRL}. During this initial decay, energy exchange between small scales and large scales still occurs through nonlinear transfer, thus maintaining the SE regime during that period.\\ (ii) Then, the final decay starts when nonlinearities become negligible, such that the viscous dissipation dominates the decay dynamics at all scales. 

Due to the time resolution limitation of the time-frequency analysis in Sect~\ref{sec:lawDiss}, we can not accurately characterize the initial decay. Thus, hereafter, we focus only on the final decay which starts at time $t_\nu$, larger than $t_0$. Experimentally, $t_\nu$ is the time at which the spectrum departs from the SE prediction of Eq.~\eqref{eq:SE_omega}.

In the final decay, Eq.~\eqref{eq:balanceEfull} then reads
\begin{equation}
\frac{\partial E(\omega,t)}{\partial t} =-D(\omega,t)\equiv -\frac{2}{\tau(\omega)}E(\omega,t),
\label{eq:balanceE}
\end{equation}
where $\tau(\omega)$ is the typical linear dissipation time scale. 
The term on the right-hand side thus corresponds to energy dissipation induced by a linear mechanism. The factor 2 comes from the definition of $\tau$ from the exponential decay {$\xi(t) \equiv \xi_\nu e^{-(t-t_\nu)/\tau}$ with $\xi_\nu\equiv \xi(t=t_\nu)$ as $t_\nu$} is the {final decay starting time}. By integrating Eq.~(\ref{eq:balanceE}) between {$t_\nu$} and $t$, the energy decays exponentially as
\begin{equation}
	E(\omega,t)= E(\omega,t_\nu)e^{-\frac{2}{\tau(\omega)}(t-t_\nu)}.
\label{eq:Eomega}
\end{equation}

As long as the final decay is not reached ($t \leq t_\nu$), we assume that the energy spectrum of the large scales follows a frequency power law. Thus, at time $t=t_\nu$, one assumes $E(\omega,t_\nu) = A\omega^\alpha$ with $A$ a constant. The exponent $\alpha$ is kept for the sake of generality, but one can bear in mind that, from Eq.~\eqref{eq:SE_omega}, $\alpha=1/3$ and $A\equiv E_0/[3\pi (\rho T^2)^{1/3}]$ in our case. Inserting $E(\omega,t_\nu)$ into Eq.~(\ref{eq:Eomega}), and the result into the total energy at time $t$,  $\cE^\omega_\mathrm{th}(t) \equiv \int_0^{\omega_c} E(\omega,t)d\omega$, yields
\begin{equation}
	\cE^\omega_\mathrm{th}(t)=\int_0^{\omega_c} A \omega^\alpha e^{-\frac{2}{\tau(\omega)}(t-t_\nu)}d\omega  \ \ \mathrm{for}\  t \geqslant t_\nu.
\label{eq:IntE}
\end{equation}
The superscript indicates whether total energy is computed from the frequency spectrum ($\cE^\omega$) or the spatial spectrum ($\cE^k$) afterwards, while the index ``$\mathrm{th}$'' indicates a theoretical expression. $\omega_c$ is a high-frequency cutoff that prevents the divergence of the integral at $t=t_\nu$ and which appears naturally since SE is observed for frequencies smaller than forcing frequencies $f_p$.

Let us assume that $\tau(\omega)$ is a power-law function of $\omega$ (as justified afterwards) as $\tau(\omega)=2/(C\omega^{\beta})$ where the exponent $\beta$ and the positive constant $C$ depend on the dissipation mechanism (see Sect.~\ref{sec:lawDiss}). The integral of Eq.~(\ref{eq:IntE}) is convergent if $\alpha>-1$ to ensure the convergence when $\omega$ tends to zero. Using the change of variable $\omega$ by $C(t-t_\nu)\omega^{\beta}$ in Eq.~\eqref{eq:IntE}, we obtain a prediction for the decay of the total energy as
\begin{equation}
	\cE^\omega_\mathrm{th}(t)=\tilde{A}(t-t_\nu)^{-\delta}\gamma[\delta,(t-t_\nu)/\tau_c],  
	\label{eq:E_total}
\end{equation}
where $\delta = (\alpha +1)/\beta$ is the time power-law exponent, $\tilde{A}=AC^{-\delta}/\beta$ is a constant, $\tau_c=\tau(\omega_c)/2$, and $\gamma$ is the lower incomplete Gamma function, $\gamma[s,x] \equiv \int_0^x u^{s-1}e^{-u}du$. When $t\rightarrow t_\nu$, Taylor's expansion of $\gamma$ scales as $\gamma[\delta, (t-t_\nu)/\tau_c]\propto (t-t_\nu)^\delta$ and thus regularizes Eq.~(\ref{eq:E_total}), i.e., prevents the divergence of $\cE^\omega_\mathrm{th}(t)$. When $x\rightarrow +\infty$, $\gamma[s,x]$ tends to $\Gamma(s)$, the usual Euler's Gamma function. The error between the two functions, $\gamma[s,x]$ and $\Gamma(s)$, is of the order $O(x^{s-1}e^{-x})$. Thus, $\cE^\omega_\mathrm{th}(t)$ is predicted to follow a time power law for $(t-t_\nu)>\tau_c$, as
\begin{equation}
	\cE^\omega_\mathrm{th}(t)\simeq\tilde{A}(t-t_\nu)^{-\delta}\Gamma(\delta).  
	\label{eq:E_total_b}
\end{equation}
This approximation will be discussed later in the light of the experimental values of $\tau_c$.

Thus, although all energy modes $E(\omega,t)$ follow exponential decays of Eq.~\eqref{eq:Eomega}, the total energy $\cE^\omega_\mathrm{th}(t)$ decreases accordingly to a time power-law of Eq.~\eqref{eq:E_total_b}. 

This result appears to be quite general and extends beyond the scope of hydrodynamic turbulence and wave turbulence. Indeed, if the energy spectrum of a system, initially follows a frequency power-law, and decays over time accordingly to a relation analog to Eq.~(\ref{eq:balanceE}), then the total energy will decay as a time power-law as Eq.~\eqref{eq:E_total_b} whose the exponent $\delta$ can be predicted as long as the linear time scale $\tau(\omega)$ is a frequency power-law function. 

In our specific case of hydroelastic tensional waves where large scales are initially in a statistical equilibrium regime, with a spectrum as in Eq.~\eqref{eq:SE_omega}, one has $\alpha=1/3$ and $A\equiv E_0/[3\pi (\rho T^2)^{1/3}]$. Moreover, the linear dissipation time $\tau(\omega)$ is predicted to scale as $\tau(\omega)=2/(C\omega^{\beta})$ with $\beta=7/6$ and $C\equiv \sqrt{\nu/2}(\rho/T)^{1/3}$ [see Sect.~\ref{sec:lawDiss} and Eq.~(\ref{eq:tau})]. Therefore, the prediction for the total energy decay reads, from Eq.~\eqref{eq:E_total_b},
\begin{equation}
	\cE^\omega_\mathrm{th}(t)\simeq\frac{2^{11/7}\Gamma(8/7)\ E_0}{7\pi (\nu^{4}\rho^{5} T^{2})^{1/7}} (t-t_\nu)^{-8/7},  	
	\label{eq:E_hydroelas}
\end{equation}
that is $\delta = (\alpha +1)/\beta=8/7$ and $\tilde{A}=AC^{-\delta}/\beta=2^{11/7} E_0/[7\pi \rho^{5/7} T^{2/7} \nu^{4/7}]$ using previous notations. 

Before testing experimentally in Sect.~\ref{sec:Decay} the large-scale energy decay prediction of Eq.~\eqref{eq:E_hydroelas}, the viscous dissipation nature of hydroelastic tensional waves is first investigated below (Sect.~\ref{sec:lawDiss}), as it prescribes the $\beta$ exponent value.

\subsection{Viscous dissipation time scale}\label{sec:lawDiss}
To obtain the exponent $\beta$, one must explore the dissipation mechanism in the system. For a plate in contact with a liquid, the boundary condition of zero tangential displacement can be obtained in two limits. In the small wave-amplitude limit, the sheet thickness is greater than the displacements of points in the sheet, which guarantees the existence of a neutral surface. On that surface, there is no compression or extension, and the horizontal displacements vanish at first order~(\S11 of \cite{Landau}). This condition generally holds for an ice layer covering the ocean~\citep{Squire2009}. Here, the wave amplitude is of the order of the typical sheet thickness $e=0.5$~mm (see bottom of figure~\ref{fig:setup}). However, applying an external pressure is crucial as the stress tensor is dominated by the constant external stretching forces~(\S14 of \cite{Landau}). In consequence, no dynamical horizontal displacement of the sheet is also expected here, and we thus assume a zero tangential velocity boundary condition at the sheet/liquid interface. The derivation of the dissipation time scale is then fully analogous to the deep-water surface wave case. In that case, the air/liquid surface boundary layer is due to an inextensible film~\citep{lamb1932,miles1967}, with $\tau^{-1}(\omega) = k\sqrt{\omega\nu}/(2\sqrt{2})$, and $\omega(k)$ given by the gravity-capillary wave dispersion relation. For hydroelastic tensional waves, $\omega(k)$ is given by the second term of the right-hand side of Eq.~\eqref{eq:LDR}, and the same procedure yields a similar expression for $\tau(\omega)$ as 
\begin{equation}
	\tau^{-1}(\omega) =  \frac{\nu^{1/2}}{2\sqrt{2}}\left(\frac{\rho}{T}\right)^{1/3}\omega^{7/6}.
	\label{eq:tau}
\end{equation}
and thus to the exponent $\beta = 7/6$. 
\begin{figure}
		\centering
		\includegraphics[width=0.7\textwidth]{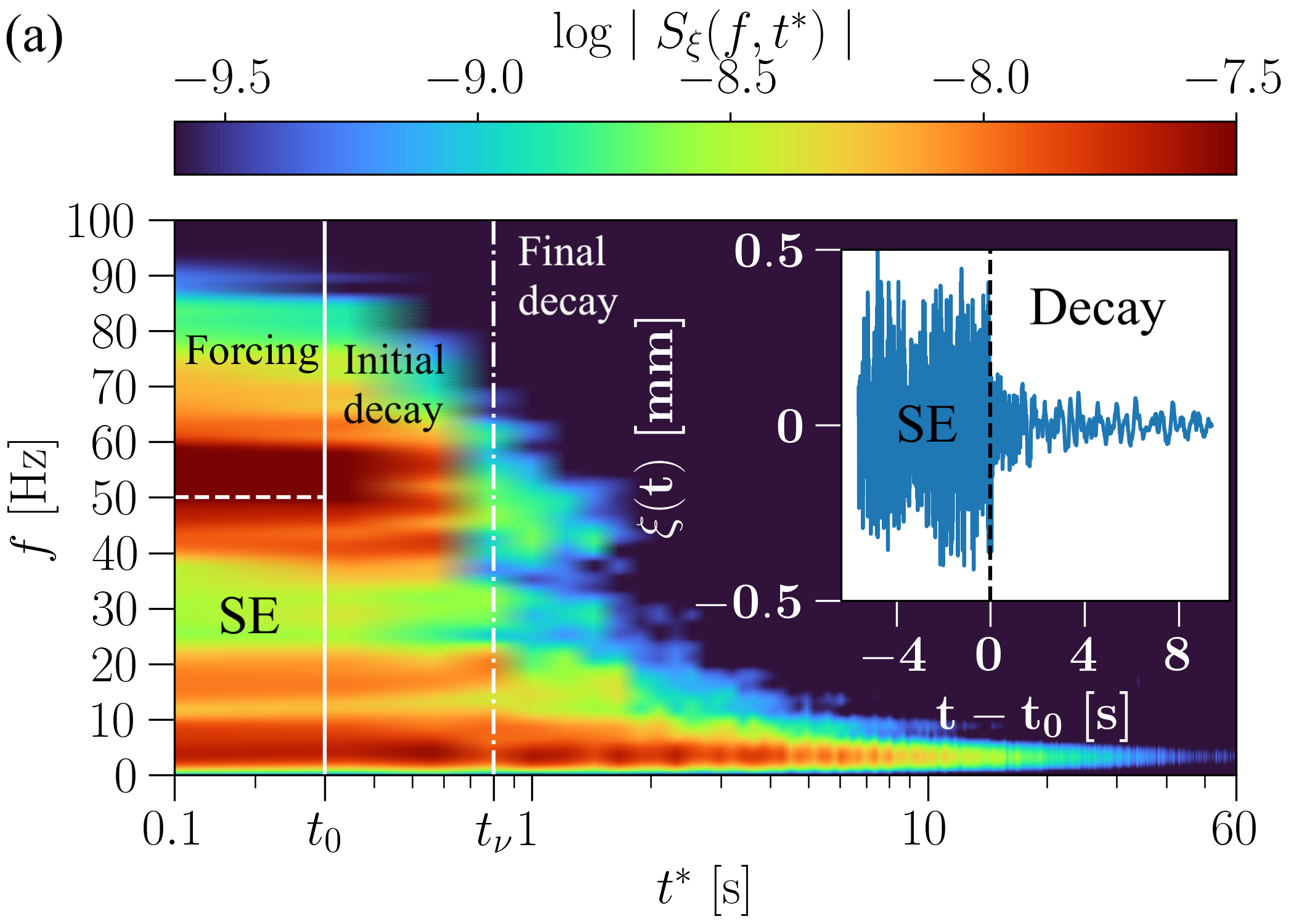}\\ 
		\includegraphics[width=0.7\textwidth]{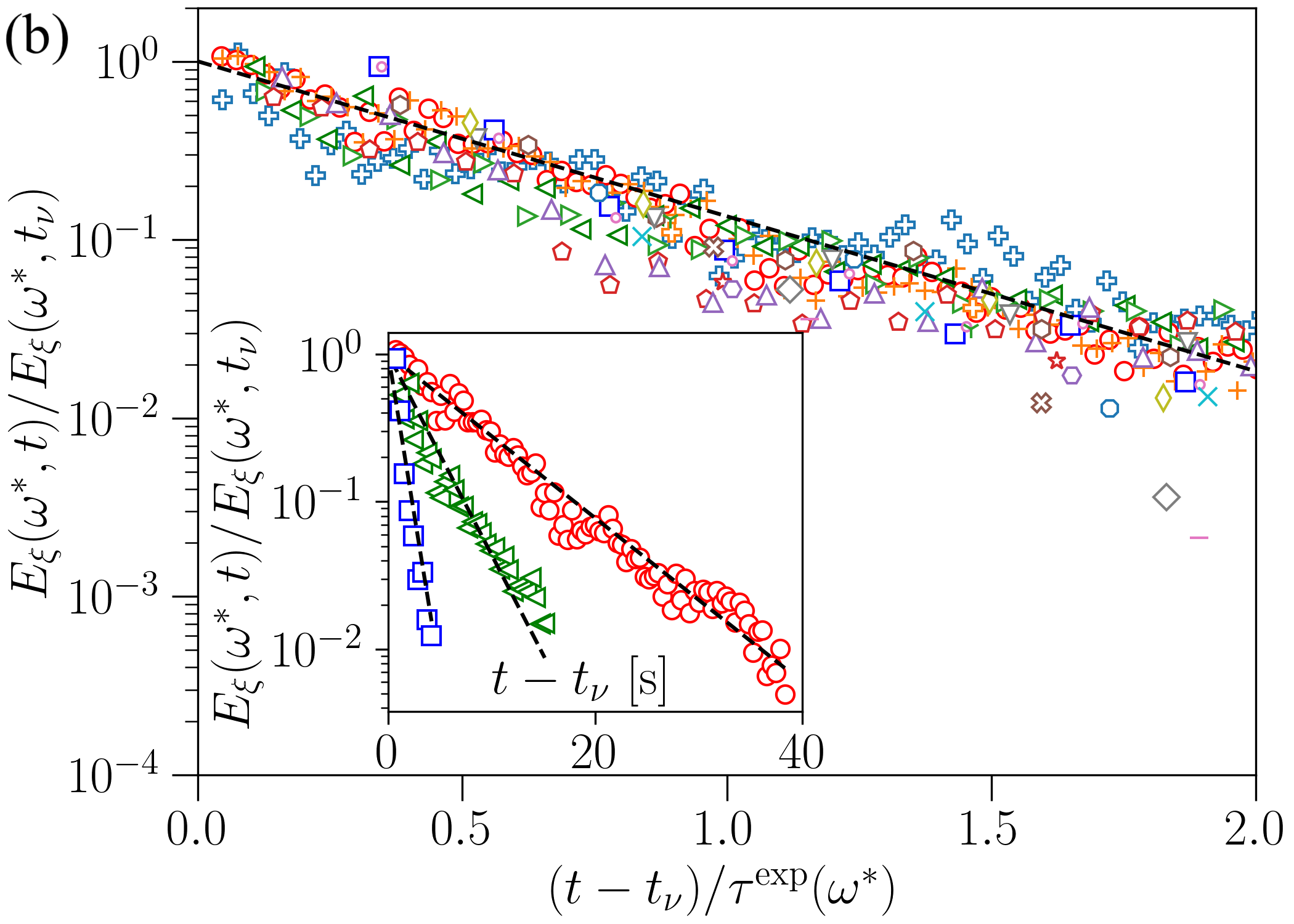} 
\caption{\label{fig:spectrogram} (a) Spectrogram $S_\xi(f,t^*)$ of the wave amplitude versus time $t^*$ (log-scale) and frequency $f$. Log-scale colorbar.  For $t<t_0$: small-scale forcing  ($f_p\in[50,100]$~Hz) and SE regime at large scales [$f< \min(f_p)$]. Forcing is stopped at time $t_0$ (vertical white solid line). For $t_0\leq t < t_\nu$: initial decaying regime. Final decay starts at time $t_\nu$ (vertical white dot-dashed line). Applied tension $T=6.8$~N~m$^{-1}$. Inset: temporal decay of wave amplitude $\xi(t)$ before ($t<t_0$) and after ($t> t_0$) forcing stops.  (b)~Normalized energy spectrum versus the normalized time $(t-t_\nu)/\tau^{\mathrm{exp}}(\omega^*)$ for various Fourier modes $\omega^*/(2\pi)\in[2,70]$~Hz (symbols). The black dashed line has a $-2$ slope corresponding to the exponential decay of Eq.~\eqref{eq:Eomega}. Applied tension $T=5$~N~m$^{-1}$. Inset: Normalized energy spectrum versus unrescaled time $t-t_\nu$ for Fourier modes $\omega^*/(2\pi)=$ 3.2($\circ$), 6.4 ($\lhd$), and 20 ($\Box$)~Hz.}
\end{figure}
 
Let us experimentally check the prediction of Eq.~\eqref{eq:tau} for the frequency-power law of the dissipation time scale, $\tau(\omega)$, and the corresponding exponential decay of the energy spectral density $E_\xi(\omega,t)$ of Eq.~\eqref{eq:Eomega}. $E_\xi(\omega,t)$ is obtained by performing a time-frequency analysis, i.e., a spectrogram of the temporal signal of the pointwise vertical position $\xi(t)$ (see inset of figure~\ref{fig:spectrogram}a) measured by the vibrometer. The spectrogram is computed as the power-spectral density at each time $t^*$
\begin{equation}
	S_\xi(\omega,t^*) \equiv \int_{t^*}^{t^*+\delta t}\langle \xi(t)\xi(t+s) \rangle_te^{-i\omega s}ds
\end{equation}
over a short time scale $\delta t = 0.5$~s. Figure~\ref{fig:spectrogram}a shows a typical example of a spectrogram, $S_\xi(\omega,t^*)$, averaged over $60$ iterations. The horizontal white dashed line corresponds to the lower bound of the forcing frequency range, $\min(f_p)$. The vertical solid line at $t^*=t_0$ indicates the time at which the external forcing is turned off. When the forcing is on ($t^*<t_0$), the energy is well spread over all frequencies lower than the forcing ones $f<f_p$ (as in the inset of figure~\ref{fig:LDR}) where SE is commonly established~\citep{vernet25}. After external forcing is switched off and final decay is initiated ($t^*>t_\nu$), all modes decay at different rates that depend on their frequency. 

The decay of the energy spectrum is then experimentally inferred from the link between the energy spectrum and the wave-amplitude power spectrum as $E_\xi(\omega,t)=(T/\rho)^{1/3}\omega^{4/3}S_\xi(\omega,t)$ (see Sect.~\ref{sec:StatEq}). The inset of figure~\ref{fig:spectrogram}b shows that the energy spectrum $E_\xi(\omega^*,t)$ decays exponentially over time, and that its decay rate, $1/\tau^{\mathrm{exp}}(\omega^*)$ (i.e., the dashed-line slope), depends on the frequency mode $\omega^*$. Each Fourier mode $\omega^*$ corresponds to a horizontal line in the spectrogram in figure~\ref{fig:spectrogram}a. When the energy spectrum $E_\xi(\omega^*,t)$, normalized by its value at $t=t_\nu$, $E_\xi(\omega^*,t_\nu)$, is plotted as a function of a normalized time $(t-t_\nu)/\tau^{\mathrm{exp}}(\omega^*)$, all data for various modes $\omega^*$ collapse on a master curve, as shown in the main figure~\ref{fig:spectrogram}b.  A clear exponential decay is observed up to $(t-t_\nu)/\tau^{\mathrm{exp}}(\omega^*)\simeq 2$ showing thus that  Eq.~\eqref{eq:Eomega} is validated experimentally. The dissipation time scale can thus be directly obtained from Eq.~\eqref{eq:Eomega} as $\tau^{\mathrm{exp}}(\omega^*)= 2\int_0^{+\infty}E_\xi(\omega^*,t)/E_\xi(\omega^*,t_\nu)dt$ due to the exponential nature of the decay.

\begin{figure}
\centering
\includegraphics[width=0.6\columnwidth]{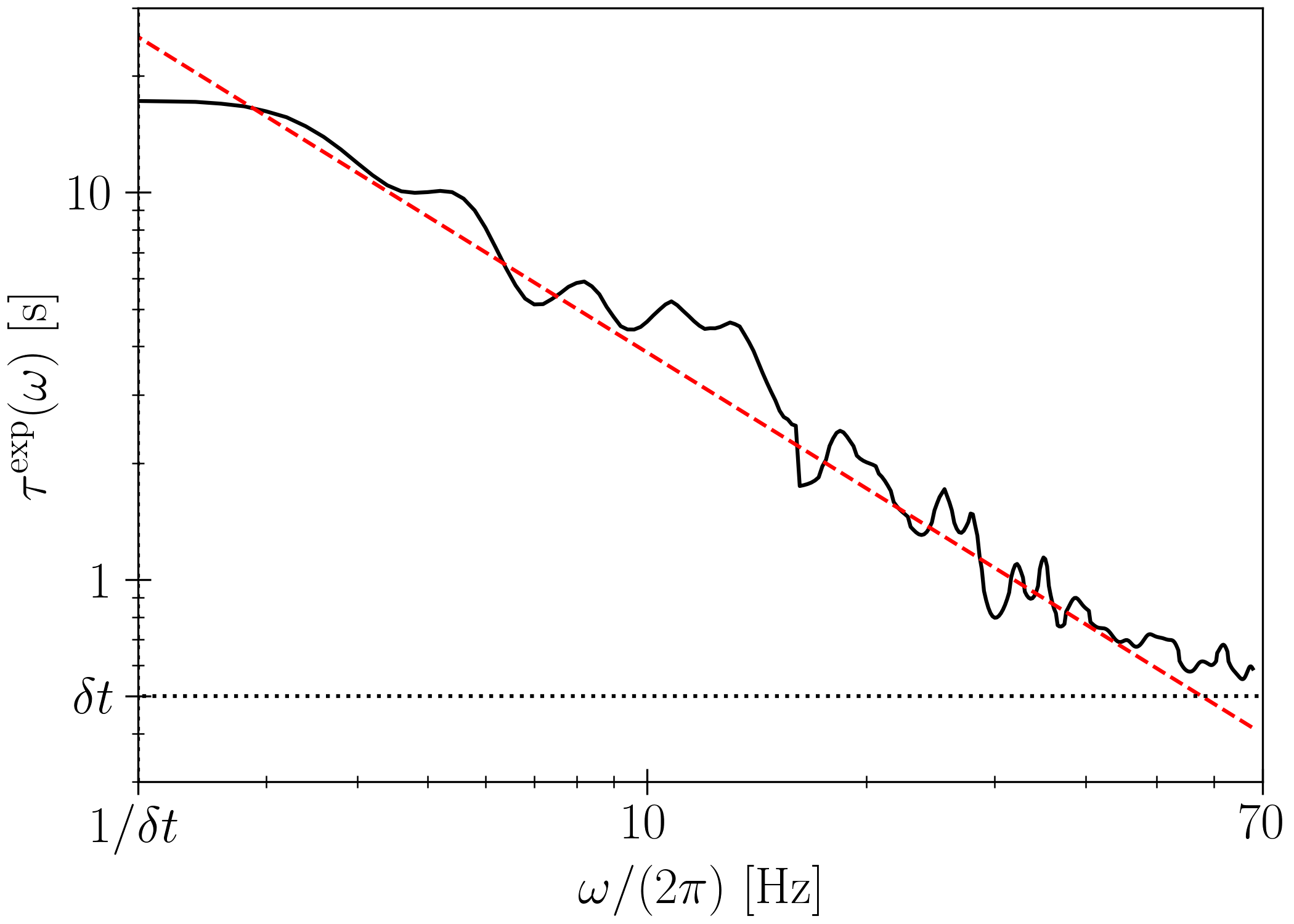} 
\caption{\label{fig:dissip_tscale} Experimental dissipation time scale $\tau^{\mathrm{exp}}(\omega)$ of hydroelastic tensional waves versus frequency $\omega/(2\pi)$. The red dashed line corresponds to Eq.~(\ref{eq:tau}) with no fitting parameter. Horizontal dotted line: temporal cutoff due to the spectrogram computation (short time interval $\delta t=0.5$~s). Cutoff frequency: $1/\delta t = 2$~Hz. Applied tension $T=5$~N~m$^{-1}$.}
\end{figure}

Figure~\ref{fig:dissip_tscale} then shows the experimental dissipation time scale as a function of the frequency. The prediction of Eq.~(\ref{eq:tau}) (red dashed line in figure~\ref{fig:dissip_tscale}) is well verified over the available frequency range with no fitting parameter. This frequency-scaling law has been checked for different tension values, but the small range of accessible tension and the weak exponent make it difficult to verify experimentally the $\tau\propto T^{1/3}$ scaling of  Eq.~(\ref{eq:tau}). As the computation relies on the spectrogram $S_\xi(\omega,t^*)$, the moving integration time $\delta t=0.5$~s is thus a lower bound for $\tau(\omega)$ as faster fluctuations are averaged. Decreasing further the value of $\delta t$ presents also two drawbacks: (i) as $S_\xi(\omega,t^*)$ converges for $f>1/\delta t$, so $\tau(\omega)$ can be computed only above this cutoff frequency, (ii) decreasing further $\delta t$ also leads to noisier spectra. This method thus estimates $\tau$ for $1/\delta t< f<60$~Hz, which is sufficient as the SE regime was initially established for frequencies lower than the forcing range $[50,100]$~Hz.

\begin{figure} 
\centering
\includegraphics[width=0.65\columnwidth]{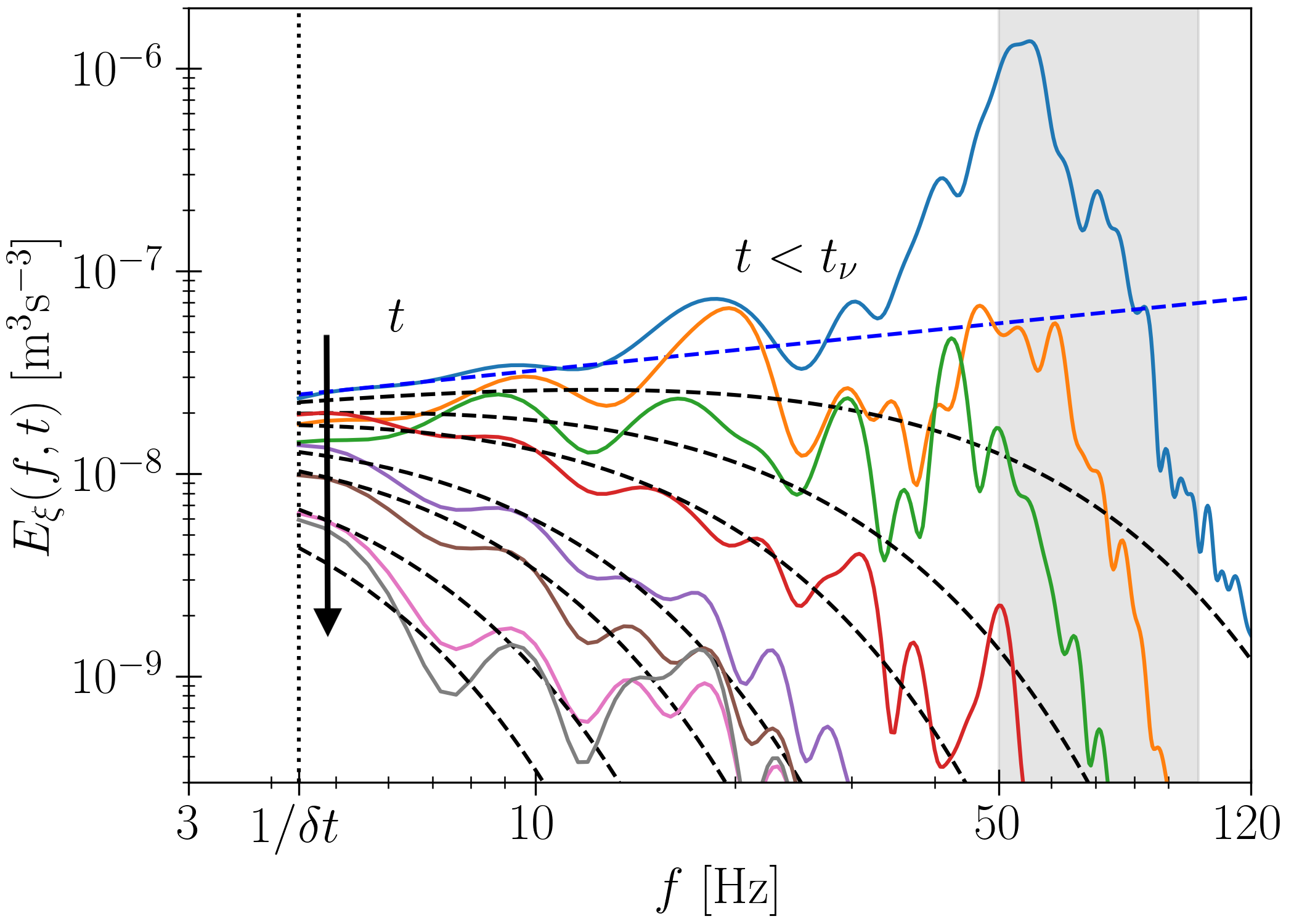} 
\caption{\label{fig:SEdecay}Energy spectrum $E_\xi(\omega,t)$ at different times, before ($t<t_\nu$ - top blue curve) and after ($t>t_\nu$ - other curves) the final decay starts. Obtained from vibrometer measurements at times $t-t_\nu \in[-0.4, 0.04,  0.7,  1.3,  2.9,  4,  6.2,  8.4]$~s (from top to bottom). The arrow indicates the direction of time. Black dashed lines correspond to Eq.~\eqref{eq:Eomega} for the same times, except the blue dashed line, for $t-t_\nu=-0.4$~s (i.e., $t-t_0=0.1$~s), corresponding to Eq.~\eqref{eq:SE_omega}. Black dotted line: cutoff frequency, $1/\delta t$, due to the spectrogram computation (short time interval of $\delta t = 0.22$~s). Grey region: initial forcing range, $f_p\in[50,100]$~Hz. Applied tension $T=6.8$~N~m$^{-1}$.}
\end{figure}

\subsection{Temporal decay of energy}\label{sec:Decay}
Figure~\ref{fig:SEdecay} shows the energy spectral density $E_\xi(\omega,t^*)$ at different decay times. Before the final decay ($t_0\leq t<t_\nu$), statistical equilibrium of large scales is observed in reasonable agreement with the stationary prediction of Eq.~(\ref{eq:SE_omega}) (top black dashed line) and the corresponding frequency power-law exponent $\alpha=1/3$. When the final decay starts ($t>t_\nu$), the stationary SE regime rapidly collapses due to spectral dissipation and the absence of energy flux. Indeed, all Fourier modes decay at different rates with $\tau^{-1}(\omega)\propto \omega^{7/6}$ [see Sect.~\ref{sec:lawDiss}, figure~\ref{fig:dissip_tscale} and Eq.~\eqref{eq:tau}], and thus modes of higher frequency decay faster than large-scale modes. Such a collapse of the spectrum is reasonably well captured by Eq.~(\ref{eq:Eomega}) (see black dashed lines in figure~\ref{fig:SEdecay}) with no fitting parameter. As discussed in Sect~\ref{sec:StatEq}, since $\alpha=1/3$ and $\beta=7/6$, the energy stored in large-scale tensional waves is predicted to decay following Eq.~(\ref{eq:E_total}) with $\delta= (\alpha +1)/\beta=8/7$, thus as Eq.~\eqref{eq:E_hydroelas} in $\cE^\omega(t)\propto (t-t_\nu)^{-8/7}$.

Such a free decay of Fourier mode energy is experimentally studied using the temporal signal {$\xi(t)$} measured at a given location by laser vibrometry. As the wave-amplitude spectrum {$S_\xi$} is related to the energy spectrum $E_\xi$ by $E_\xi(\omega,t^*)=\omega^2 S_\xi(\omega,t^*)/k$, we obtain ${\cE^\omega}(\omega,t^*)$ from the dispersion relation and the spectrogram ${S_\xi}(\omega,t^*)$ over a short-time window $\delta t=0.22$~s as
\begin{equation}
	{\cE^\omega}(t^*) \equiv \int_{2\pi/\delta t}^{{\omega_c}}\frac{T}{\rho}k^2 {S_\xi}(\omega,t^*)d\omega,
	\label{eq:Ew}
\end{equation}
with $\omega_c/(2\pi)=30$~Hz as SE is reasonably established for $t\lesssim t_\nu$ {up} to this frequency.

To increase the statistical reliability of experimental data for $\cE^\omega(t^*)$, measurements are performed over 60 decay experiments of $60$~s each. Between each decay, forcing is maintained over $60$~s long enough to recover a steady SE regime. 

\begin{figure}
	\includegraphics[width=0.5\textwidth]{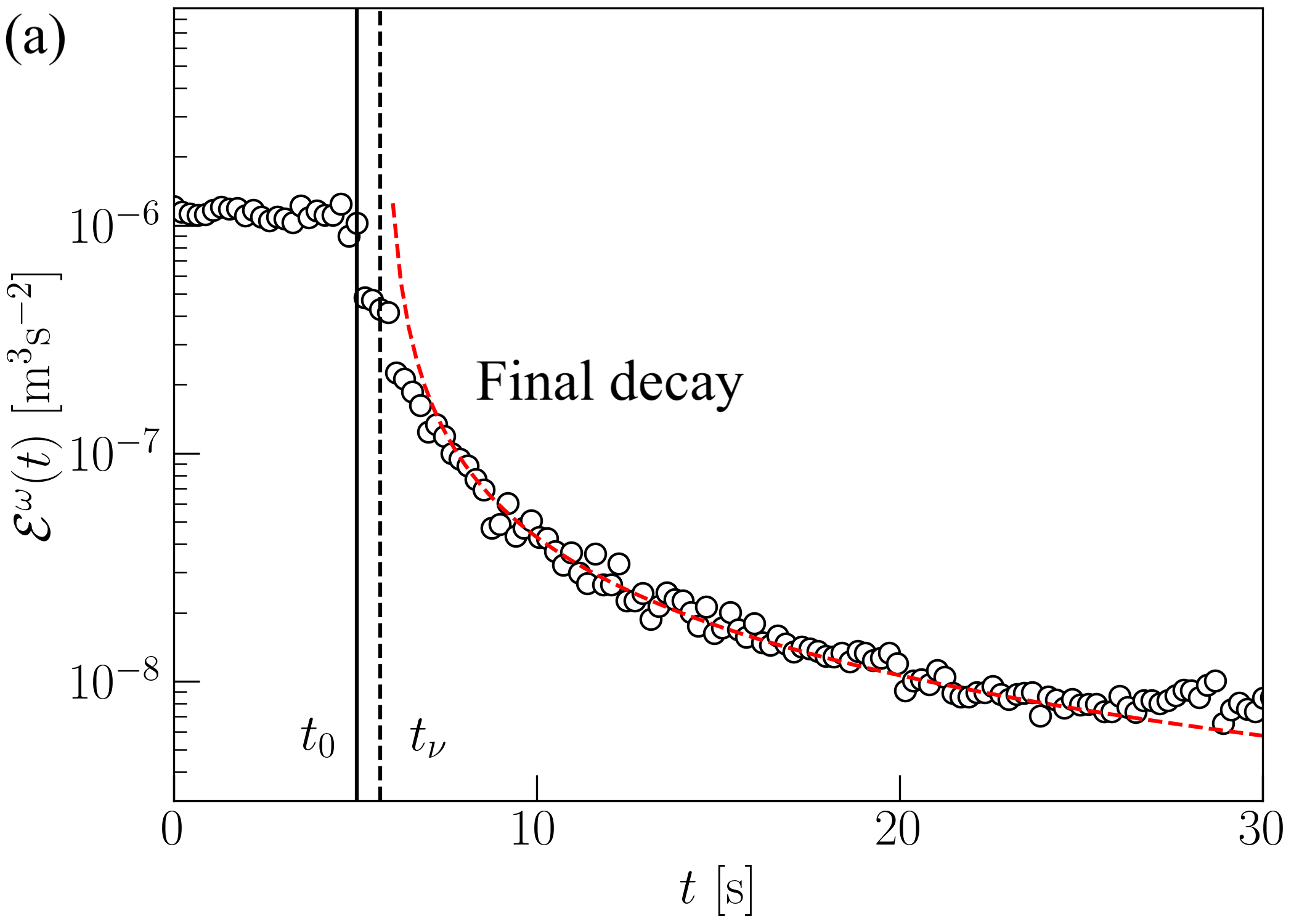}
	\includegraphics[width=0.5\textwidth]{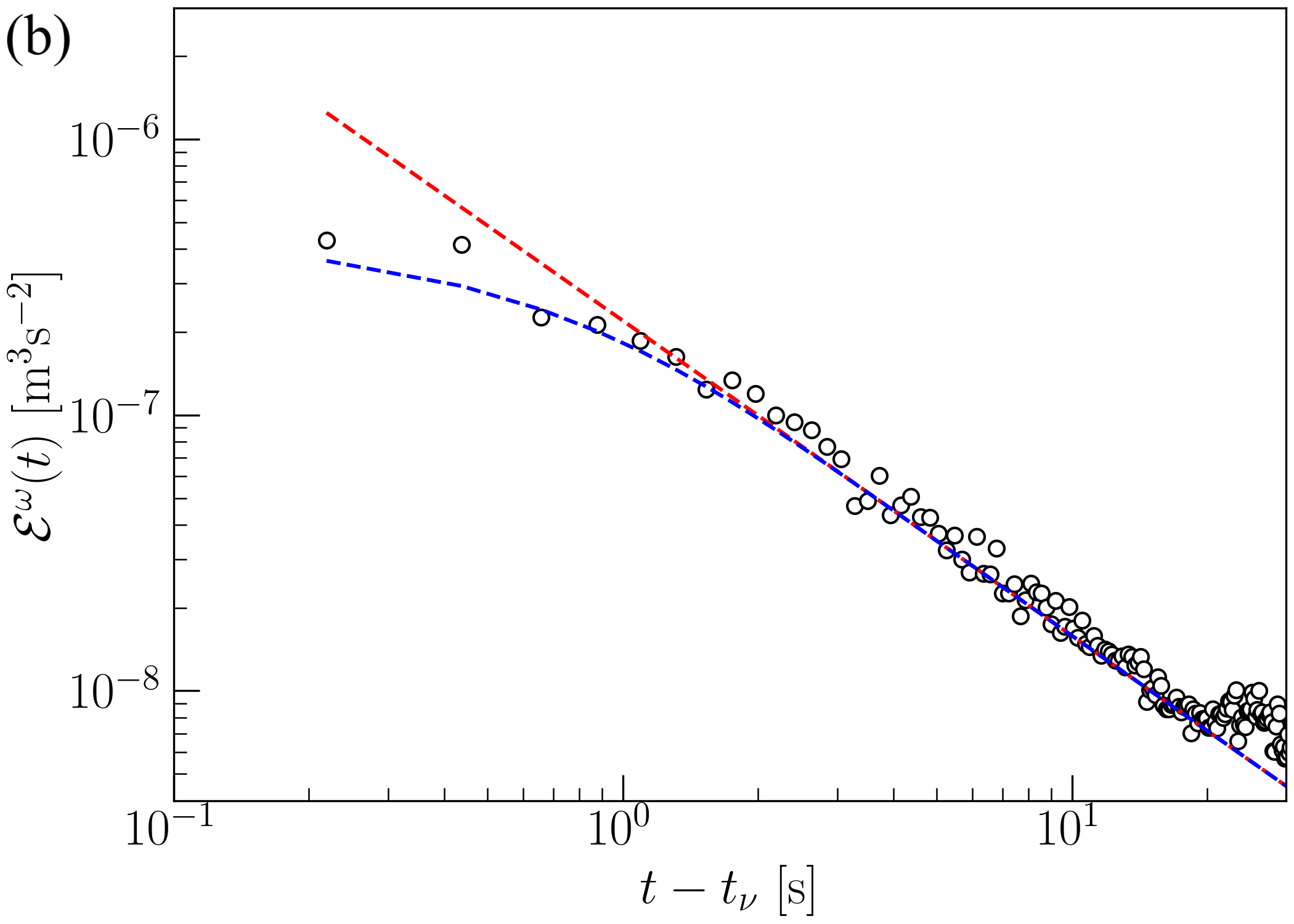}\\ 
	\includegraphics[width=0.5\textwidth]{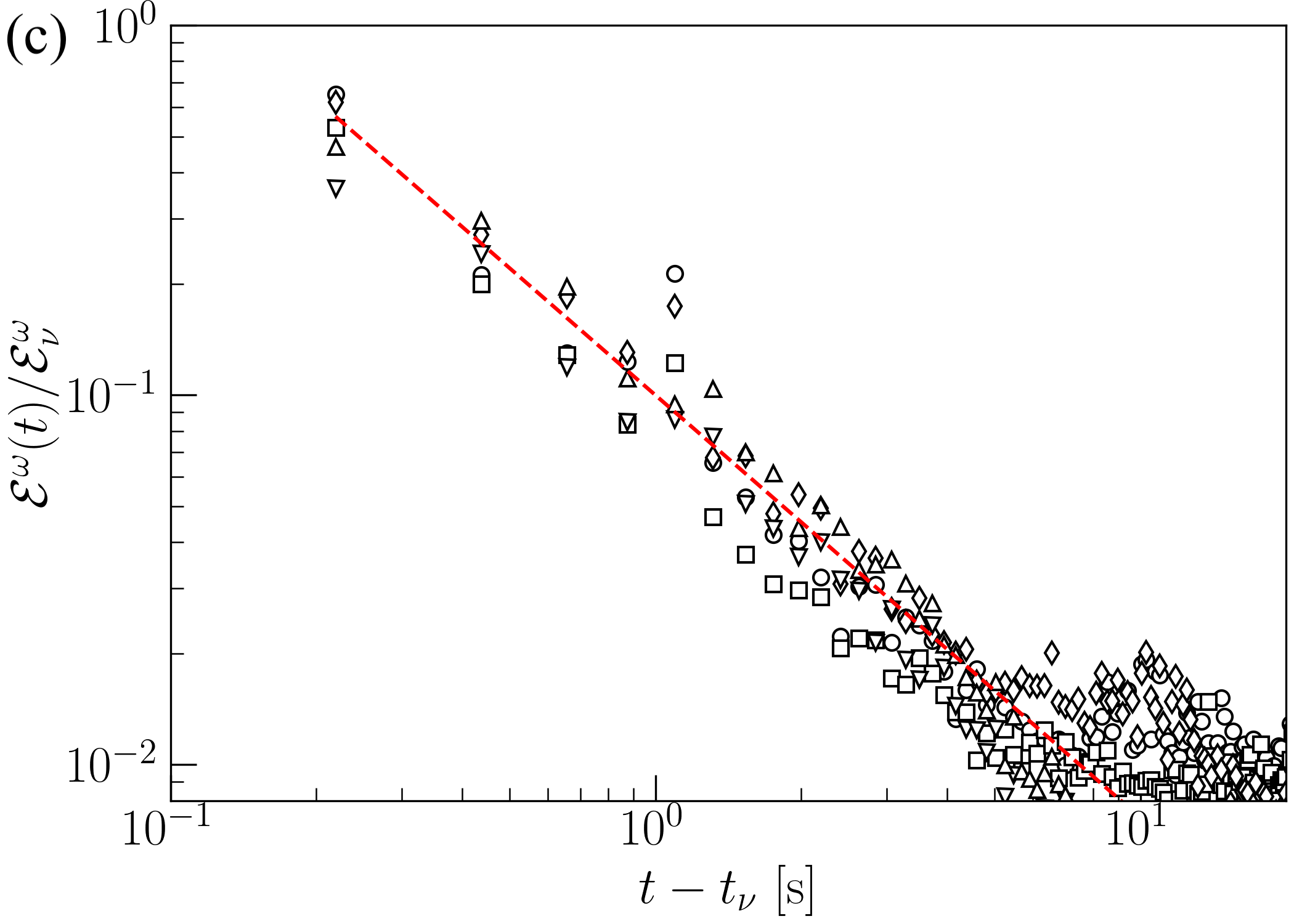} 
	\includegraphics[width=0.5\textwidth]{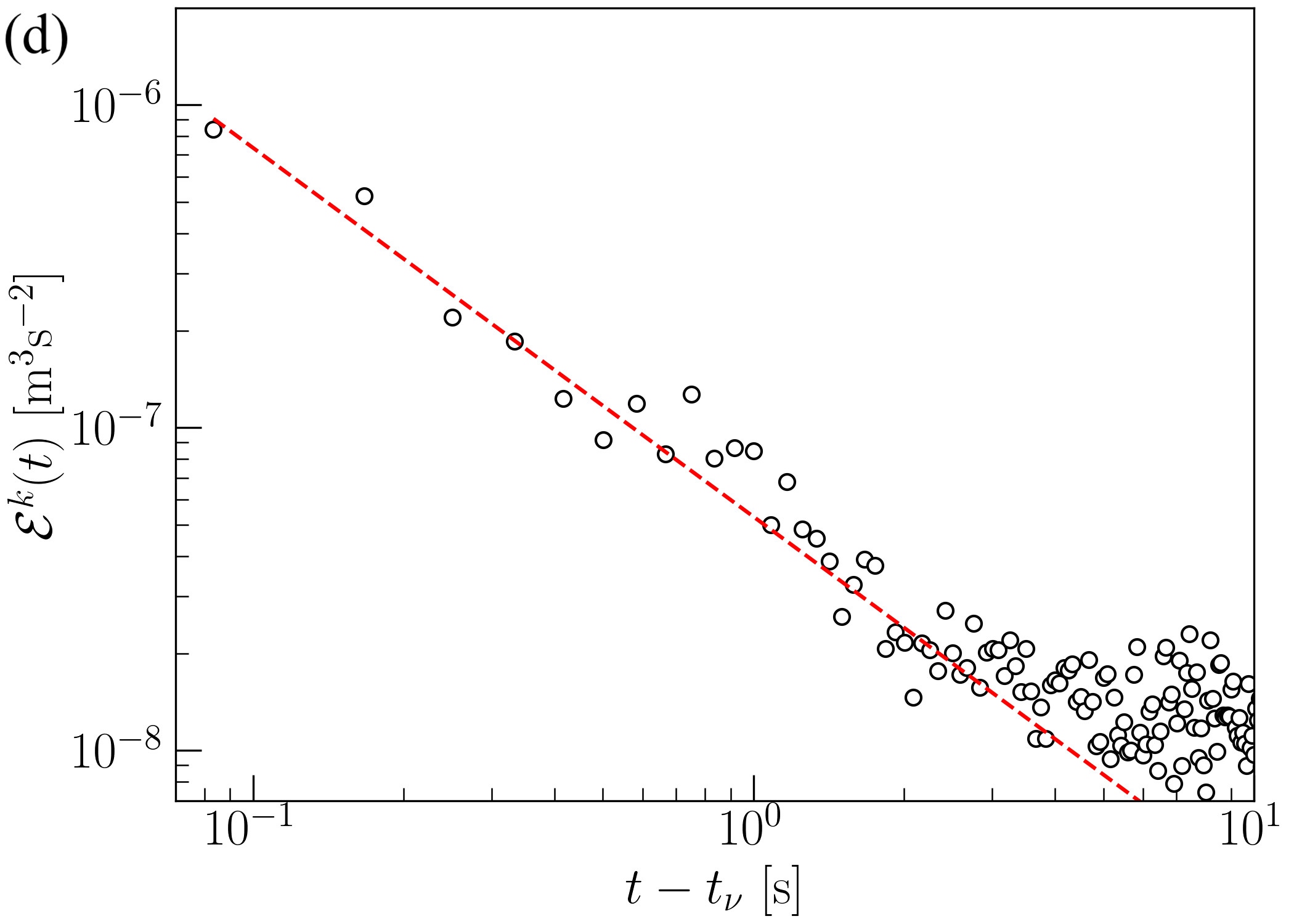}
\caption{\label{fig:temp_decay}(a) Temporal evolution (semilog-y plot) of the total energy $\cE^\omega(t)$ before ($t<t_0$) and after ($t> t_0$) forcing stops. Vertical solid line corresponds to $t=t_0$. Vertical dashed line corresponds to the start of the final decay at $t=t_\nu$. The initial decay lasts $t_\nu-t_0\simeq 0.5$~s and the final decay of the order of $20$~s.  (b) Temporal decay of the total energy $\cE^\omega(t)$, estimated by laser vibrometry and averaged over $60$ decay experiments. Red-dashed line has a $-8/7$ slope corresponding to the prediction of Eq.~\eqref{eq:E_hydroelas}. Full prediction of Eq.~\eqref{eq:E_total} corresponds to the blue-dashed line. Applied tension: $T=6.8$~N~m$^{-1}$. (c) $\cE^\omega(t)$ rescaled by its initial value for various initial forcing strengths $E_0 \in [1.5, 15]\times 10^{-8}$~m$^3$.s$^{-2}$ (from $\circ$- to $\triangle$-symbols). Red-dashed line has a $-8/7$ slope [Eq.~\eqref{eq:E_hydroelas}]. (d)~Temporal decay of the total energy $\cE^k(t)$ estimated by FTP. Red-dashed line has a $-8/7$ slope [Eq.~(\ref{eq:E_hydroelas})]. Applied tension $T=6.8$~N~m$^{-1}$. }
\end{figure}

Figure~\ref{fig:temp_decay}a shows the temporal evolution of the total energy $\cE^\omega(t)$, computed from Eq.~\eqref{eq:Ew}, slightly before stopping forcing ($t<t_0$), where a steady SE regime occurs, and, after stopping ($t>t_0$), where the decay is observed during 30~s. Before collapsing, the SE prediction is maintained over a short period of time ($t_\nu-t_0\sim 0.5$~s, i.e., $2$ to $3\delta t$) during an initial period of decay. A final decay ($t>t_\nu$) then occurs during roughly 20~s. 
The main figure~\ref{fig:temp_decay}b then shows the temporal decay of $\cE^\omega(t)$ in a log-log plot. $\cE^\omega(t)$ shows a time power-law decay well described, over almost two decades, by the prediction of Eq.~\eqref{eq:E_hydroelas} in $(t-t_\nu)^{-8/7}$ (red-dashed line), up to $t-t_\nu \sim 20$~s. Note this self-similar decay does not rely on a self-similar spectrum during the decay (see figure~\ref{fig:SEdecay}). Similar results have been obtained at different measurement locations and for different applied tensions $T\in[3,7]$~N~m$^{-1}$. Moreover, the $-8/7$ exponent is independent of the strength of the initially applied forcing, thus for different energies $\cE^\omega(t=t_\nu)$ of the initial SE regime (see figure~\ref{fig:temp_decay}c), as expected by Eq.~\eqref{eq:E_hydroelas}. Note that only the total duration of the decay is modified. As detailed in Sect.~\ref{sec:cool}, the time power-law prediction of Eq.~\eqref{eq:E_hydroelas} is accurate only for $t-t_\nu>\tau_c$ with $\tau_c\sim 0.3-0.5$~s. {As shown in figure~\ref{fig:temp_decay}b, the full prediction of Eq.~\eqref{eq:E_total} (blue-dashed line) departs from this time power law} only for $(t-t_\nu)/\tau_c\lesssim 2$, i.e., $(t-t_\nu)\lesssim 1$~s. 

Finally, let us use the spatiotemporal measurements of $\eta(x,y,t)$ obtained by FTP to probe the energy decay in the spatial Fourier space $\cE^k(t)$. We expect that the energy decay, computed with the two different methods, will be the same, i.e., $\cE^\omega(t)=\cE^k(t)$. The total energy $\cE^k(t)$ is then defined by
\begin{align}
	\cE^k(t) = \int_{k_\mathcal{L}}^{k_c}\frac{T}{\rho}k^2 \langle \tilde{S}_\eta(k,t)\rangle_{\delta\tau} dk,
	\label{eq:Ek}
\end{align}
where $k_c=k(\omega_c)$, $k_\mathcal{L}=2\pi/\mathcal{L}$, and $\tilde{S}_\eta(k,t) = \langle \vert{ \check{\eta}}(k,\varphi,t) \vert^2 \rangle_\varphi/\mathcal{L}^2$ is the spatial power spectrum of the wave amplitude, $\check{\eta}(k,\varphi,t)$ is the spatial Fourier transform of $\eta(x,y,t)$, and $\langle \cdot \rangle_{\delta\tau}$ denotes a temporal average over a short time window $\delta\tau=8.3\times 10^{-2}$~s to slightly smooth the spectrum fluctuations. Figure~\ref{fig:temp_decay}d then shows the temporal decay of the total energy, $\cE^k(t)$, computed from Eq.~\eqref{eq:Ek} over the spatial scales fulfilling SE, i.e., from $k_\mathcal{L}$ to $k_c$. Here also, $\cE^k(t)$ shows a power-law decay well described, over almost two decades, by the prediction of Eq.~\eqref{eq:E_hydroelas} in $(t-t_\nu)^{-8/7}$ (red-dashed line). After a typical time $t-t_\nu \sim 4$~s, the spatiotemporal measurement reaches a signal-to-noise ratio close to unity. The duration of decay ($\sim 20$~s) is longer using local vibrometry measurements due to its better vertical resolution. Overall, both methods converge to the same decaying exponent $-8/7$ predicted by Eq.~\eqref{eq:E_hydroelas} during the final decay. 
 
\section{Conclusions}
We experimentally reported on the free decay of large-scale hydroelastic turbulent waves, initially in a state of statistical equilibrium (SE) where energy is equipartitioned among large-scale modes. Using space- and time-resolved measurements, we showed that the total energy of such large-scale tensional waves decays over time as $\cE(t)\propto (t-t_\nu)^{-8/7}$, where $t_\nu$ denotes the start of the final decay. We derived the energy decay law from the theoretical SE spectrum and linear viscous damping. This prediction is in good agreement with experimental data over nearly two decades in time, for various initial energy inputs of the SE state. We also identified the dissipation mechanism as arising from a zero horizontal velocity boundary condition at the sheet-liquid interface, and confirmed the predicted dissipation rate $\tau(\omega)$ experimentally over more than a decade in frequency. To our knowledge, this represents the first experimental validations of both $\cE(t)$ and $\tau(\omega)$ for hydroelastic tensional waves. Our approach is broadly applicable to other decaying turbulence or wave-turbulence systems initially in a large-scale SE regime. 

Once forcing is stopped, a short initial decay driven by nonlinearities occurs before a long viscous final decay initiates. Within this final decay, each energy Fourier mode decays over a different timescale, leading to a rapid breakdown of the SE regime as high-frequency (i.e., small-scale) modes dissipate faster. This collapse of SE prevents the definition of effective temperature and entropy, which are only meaningful in the stationary regime~\citep{vernet25}. Although weak dissipation allows SE to form under steady forcing, it dominates once the final decay starts, making statistical mechanics inapplicable. Still, the energy decay retains a memory of the initial SE through its influence on the time power-law decay exponent of the total energy. This behavior contrasts with classical wave turbulence, where energy stored in the largest scale continues to cascade toward smaller scales during decay, preserving a self-similar spectrum at early time~\citep{deike2012,cazaubiel2019}. The scenario described here is thus the wave turbulence counterpart of the free decay of large-scale SE in 3D turbulence reported by~\cite{gorce2024PRL}.

Although our experimental system focuses on tensional waves, i.e., hydroelastic waves that differ from the bending-dominated regime relevant to ice floes, these two systems share several common properties. Both tensional and bending waves exhibit three-wave resonant interactions, which thus inhibit an inverse cascade of energy, making the SE regime relevant for large-scale turbulent bending waves. Moreover, the dissipation mechanism identified here should also apply to the sea-ice case, although whether it is the dominant pathway for energy dissipation remains unclear. This question is still open in geophysics and lies beyond the scope of the present study.

Finally, several questions remain open when considering nonstationary processes. Namely, what is the mechanism behind the emergence of SE? What is the SE dynamics when the external random forcing is slowly modulated in time? From a broader perspective, and for the steady SE regime, how far can the analogy to statistical mechanics be pushed? For instance, is there a possibility for the Fluctuation Theorem and the Fluctuation-Dissipation relation to hold in these systems in SE? Answering these questions using statistical mechanics concepts would undoubtedly strengthen our understanding of the large scales of turbulent systems in SE.

\section*{Acknowledgement} We thank M. Lanoy and S. Fauve for fruitful discussions, and Y. Le Goas and A. Di Palma for technical support. 

\section*{Funding}This work was supported by the Simons Foundation MPS-WT-00651463 Project (U.S.) on Wave Turbulence and the French National Research Agency (ANR Sogood Project No.~ANR-21-CE30-0061-04 and ANR Lascaturb Project No.~ANR-23-CE30-0043-02).

\section*{Declaration of interests} The authors declare no conflicting interests.

\bibliographystyle{jfm}

\end{document}